\documentclass[10pt,a4paper]{article}
\usepackage[utf8]{inputenc}
\usepackage{amsmath}
\usepackage{amsfonts}
\usepackage{amssymb}
\usepackage{xfrac}
\usepackage{bm}
\usepackage{upgreek}
\usepackage{graphicx}
\usepackage[all]{nowidow}
\usepackage[left=2.5cm,right=2.5cm,top=2.5cm,bottom=2.5cm]{geometry}
\usepackage{cite}
\usepackage{dirtytalk}
\usepackage{float}
\usepackage[doublespacing]{setspace}
\usepackage{authblk}
\usepackage{setspace}
\usepackage{color,soul}
\usepackage{xcolor}

\begin{document}

\title{Utilizing the sensitization effect for direct laser writing in a novel photoresist based on the chitin monomer N\hbox{-}acetyl\hbox{-}\textsc{D}\hbox{-}glucosamine}

\author[1]{Dominic T. Meiers\thanks{contributed equally to this work}}
\author[2]{Maximilian Rothammer$^*$}
\author[2]{Maximilian Maier}
\author[2]{Cordt Zollfrank}
\author[1,3]{Georg von Freymann}

\affil[1]{Physics Department and Research Center OPTIMAS, Technische Universit\"at Kaiserslautern, 67663 Kaiserslautern, Germany}

\affil[2]{Chair for Biogenic Polymers, Technische Universität München, 94315 Straubing, Germany}

\affil[3]{Fraunhofer Institute for Industrial Mathematics ITWM, 67663 Kaiserslautern, Germany}

\maketitle
\newpage

\onehalfspacing

\begin{abstract}

The great flexibility of direct laser writing arises from the possibility to fabricate precise three-dimensional structures on very small scales as well as the broad range of applicable materials. However, there is still a vast number of promising materials which are currently inaccessible requiring the continuous development of novel photoresists. Here, a new bio-sourced resist is reported which relies on the monomeric unit of chitin, N\hbox{-}acetyl\hbox{-}\textsc{D}\hbox{-}glucosamine, expanding the existing plant-based biopolymer resists by a bio-based monomer from the animal kingdom. In addition it is shown that combined use of two photoinitiators is advantageous over the use of a single one. In our approach, the first photoinitator is a good two-photon absorber at the applied wavelength, while the second photoinitiator exhibits poor two-photon absorbtion abilities, but is better suited for crosslinking of the monomer. The first photoinitiator absorbs the light acting as a sensitizer and transfers the energy to the second initiator, which subsequently forms a radical and initializes the polymerization. This sensitization effect enables a new route to utilize reactive photointiators with a small two-photon absorption cross-section for direct laser writing without changing their chemical structure.

\end{abstract}


\section{Introduction}

Since its invention in the mid 1990s \cite{Maruo97} and its commercialization a decade afterwards direct laser writing (DLW) rapidly became the most versatile tool for fabricating three-dimensional structures from sub-micron to centimeter scales. Today, direct laser written structures find application in almost all fields requiring structures with microscopic features such as microoptics \cite{Gissibl16,Kubec22,Li22}, microfluidics \cite{AU16,Luken21,Michas22}, biomimetics \cite{Rothammer21,Wang22,Pompe22} and life science \cite{Hohmann14,Jang19,Akolawala22}. 

In contrast to other optical lithography techniques, DLW relies on the simultaneous absorption of two photons, i.e. a non-linear process, to initialize the polymerization of an appropriate photoresist. In consequence the absorption exceeds the polymerization threshold only in the tight focus of a femtosecond laser beam. This limits the polymerized area in all three dimensions and therefore enables the fabrication of almost arbitrary three-dimensional structures with feasible feature size and resolution (i.e. smallest distance between two features) below 100 nm and 500 nm, respectively, depending on the setup \cite{Hohmann15}.

In addition to optimizing the underlying technology to improve resolution and fabrication speed, the range of available materials that can be patterned by DLW has rapidly expanded over the last decade to meet different applications. Originally starting with a few polymer-based resists, today there is a whole range of such resists designed for different length scales, writing speeds, optical, mechanical or surface properties \cite{Kiefer20,Schmid19,Durisova18,Pertoldi20,Mayer20}. However, DLW-suitable photoresists are not limited to traditional polymer-based ones, but it is also possible to structure ceramics \cite{Bauer19,Farrell21}, metals \cite{Tanaka06,Waller18}, hydrogels \cite{Nishiguchi18,Hippler19}, glass \cite{Kotz21} and biomaterials \cite{Serien15,Rothammer18,Skliutas20,Grauzeliene21}. Biopolymers, which unlike conventional petroleum-based polymers are derived from renewable resources, are of particular interest for the fabrication of biomimetic structures and can serve as biocompatible materials in biomedical applications. Although animal- and plant-based biopolymers abundantly occur in nature \cite{Rao14} and both are used in various applications \cite{Sani22,Wankhade20}, biopolymer-based photoresists are currently limited to plant-based materials \cite{Rothammer18}.

Here, we present the synthesis of a novel, functionalized N\hbox{-}acetyl\hbox{-}\textsc{D}\hbox{-}glucosamine (NAG) derivative, which can be applied as monomer in a DLW-suitable photoresist. Since NAG is the monomeric unit of chitin (the building block of the exoskeletons of e.g. arthropods \cite{Rao14}), the reported resist expands the range of available bio-based photoresists to include those of animal origin. 

The NAG is functionalized through the addition of methacrylic side groups which can be crosslinked by an appropriate photoinitiator. While it is possible to structure the resist with a single photoinitiator, we demonstrate that the mixture of two different types is advantageous. Thereby, a similar principle is used as applied in pigmented UV curing, where the pigments in a lacquer show a stronger UV absorption than the photoinitiator. This shielding effect can be overcome by adding a sensitizer absorbing at a different (longer) wavelength, which is capable to transfer the absorbed energy to the photoinitiator forming highly reactive radicals \cite{Green10}. The curing speed for pigmented lacquers can be largely increased using this sensitization effect \cite{Rist92}.

In case of photoresists there is no shielding issue but a mismatch between the used wavelength and the two-photon absorption (TPA) maximum of the photoinitiator leading to a small TPA cross section (TPCS). Therefore, we mix a sensitizer possessing a TPA maximum close to the used wavelength with a photoinitiator having a maximum noticeably below the applied wavelength, but being more effective to crosslink the monomer. As reported here, such a mixture indeed outperforms resists containing either solely a sensitizer or a photoinitiator. 

\section{Results and Discussion}

\subsection{Chemical modification of N\hbox{-}acetyl\hbox{-}\textsc{D}\hbox{-}glucosamine}

To provide photo-crosslinking capabilities to the monosaccharide N\hbox{-}acetyl\hbox{-}\textsc{D}\hbox{-}glucosamine, unsaturated side groups are introduced via \textit{N,N}\hbox{-}dicyclohexylcarbodiimide-coupled esterification during the synthesis. Methacrylic acid anhydride was employed for a functionalization of the hydroxyl groups of NAG and resulted into methacrylated N\hbox{-}acetyl\hbox{-}\textsc{D}\hbox{-}glucosamine (MANAG). However, exact positions of functionalization cannot be unambiguously determined due to overlapping signals and complex patterns in the nuclear magnetic resonance (NMR) spectra. The reactive methacrylic moiety is essential for a subsequent photopolymerization reaction. 

\begin{figure}[th]
    \centering
    \includegraphics[width=\textwidth]{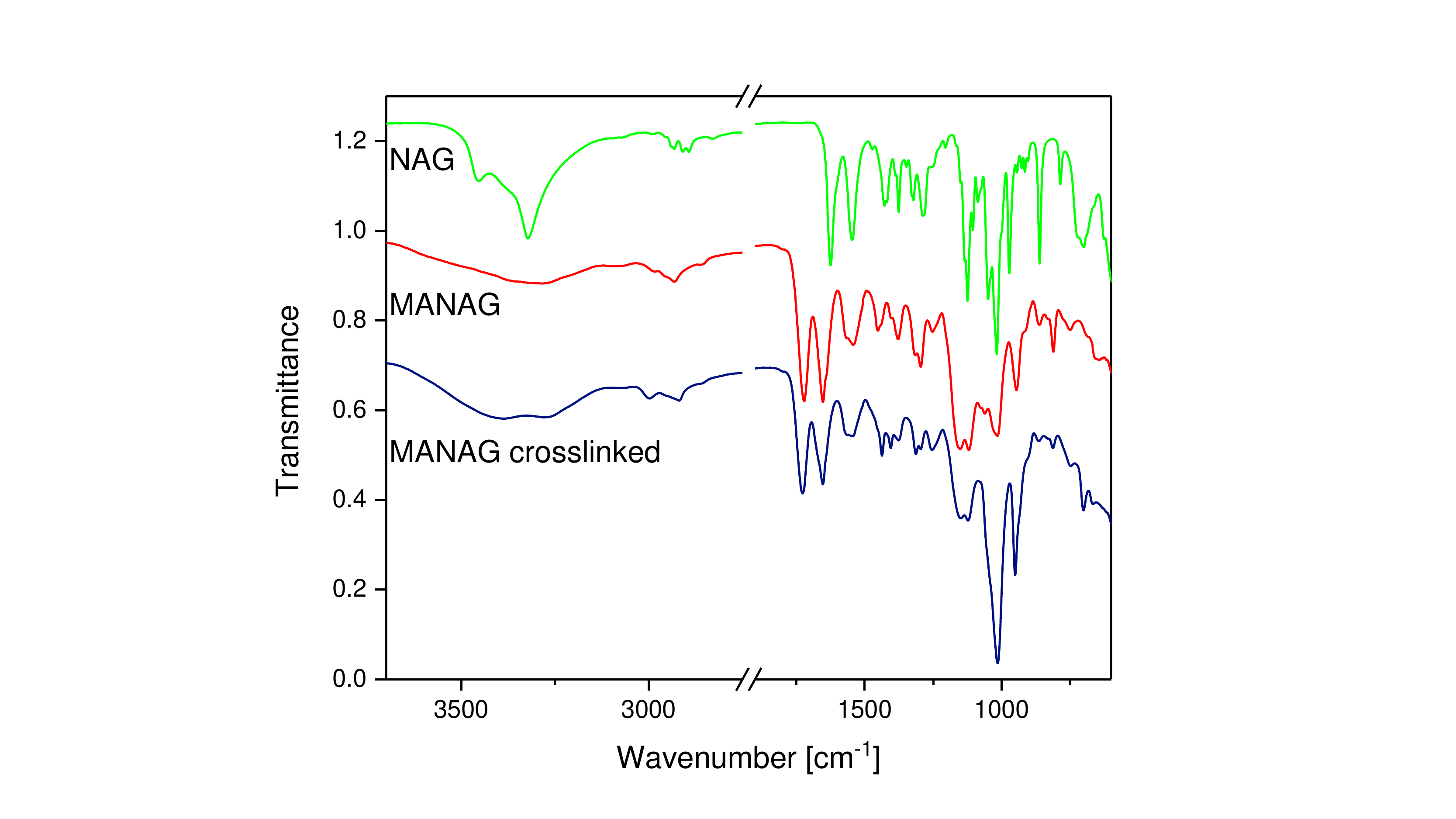}
    \caption{FTIR spectra of NAG, MANAG and crosslinked MANAG. The spectra are vertically shifted for better readability.}
    \label{fig:FTIR}
\end{figure}

Fourier transform infrared spectroscopy (FTIR) analyses of NAG and MANAG verify the successful esterification of NAG with the unsaturated anhydride. The most characteristic intensities of absorption bands of NAG are displayed in Figure \ref{fig:FTIR}. 

The absorption bands at 1624\,cm$^{-1}$ and 1546\,cm$^{-1}$ correspond to the amide\,I and amide\,II modes of the amide carbonyl group in secondary amides, respectively \cite{She1974, Belfer1998, Alkrad2003, Wang2013, Chen2015, Ji2020}. The amide I band is mainly caused by C=O stretching vibrations and by a minor ratio of N-H bending and C-H and C-N stretching vibrations of amides\cite{Bonnin1999, Sizeland2018}. The amide II band is assigned to N-H bending and some C-N stretching vibrations \cite{Bonnin1999, Sizeland2018, Du2022}. The bands at 3453\,cm$^{-1}$ and 3322\,cm$^{-1}$ represents the O-H and N-H stretching vibrations \cite{Alkrad2003, Chen2015, Cebe2020}. C-O-C antisymmetric bridge oxygen stretching vibrations, C-O bond stretching vibrations and antisymmetric in-phase ring vibrations and C-O-H stretching vibrations are located at 1125\,cm$^{-1}$, 1049\,cm$^{-1}$ and 1018\,cm$^{-1}$ \cite{Alkrad2003, Wang2013, Chen2015, Cebe2020, Du2022}. New bands arise at 1721\,cm$^{-1}$ and 812\,cm$^{-1}$ after the methacrylation reaction. The band at 1721\,cm$^{-1}$ correlates to the carbonyl C=O stretching vibrations of the ester \cite{She1974, Bonnin1999, Khor2011, Wang2013} and therefore, verifies a successful functionalization of NAG. The C=CH$_{2}$ out of plane deformation vibration of the methacrylate group appears at 812\,cm$^{-1}$ \cite{Khor2011, Rothammer18} and confirms the presence of carbon double bonds. Furthermore, a decrease of the absorption band at 3453\,cm$^{-1}$ corresponding to the hydroxyl groups is observably comparing the NAG and MANAG spectra. This confirms substitution at the OH-groups by methacrylate groups. 

Photo-crosslinking of MANAG results in a high decrease of the methacrylic carbon double bond $\nu$(C=C) intensity at 812\,cm$^{-1}$, indicating a consumption of the double bonds during polymerization of the monomers.

Moreover, the evaluation of the $^{1}$H and $^{13}$C NMR spectra also validates an effective methacrylation of NAG. The characteristic $^{13}$C NMR signals of MANAG and NAG are depicted in Table \ref{fig:NMR}. NAG is identified by the signals at 170.0\,ppm, 91.1–54.8\,ppm and 23.2\,ppm. The signals from 91.1–54.8\,ppm are assigned to the C1–C6 atoms of the glucose backbone of the monosaccharide \cite{Dang2014}. 170.0\,ppm represents the ester and 23.2\,ppm the methyl group of the N-acetate moiety, respectively \cite{Dang2014}. After the synthesis, new peaks arise at 166.9\,ppm, which reveals the esterification of NAG with methacrylic acid anhydride \cite{Dang2014, Rothammer18}. The signals corresponding to the carbon double bond are located at 136.3\,ppm and 126.3\,ppm \cite{Rothammer18}. Furthermore, the peak for the CH$_{3}$ group of the methacrylic side group is detected at 18.6\,ppm \cite{Khor2011}.

\begin{table}[th]
    \centering
     \caption{: Signal assignment for the $^{13}$C NMR spectra of NAG and MANAG, analyzed in DMSO-d$_{6}$.}
    \includegraphics[width=\textwidth]{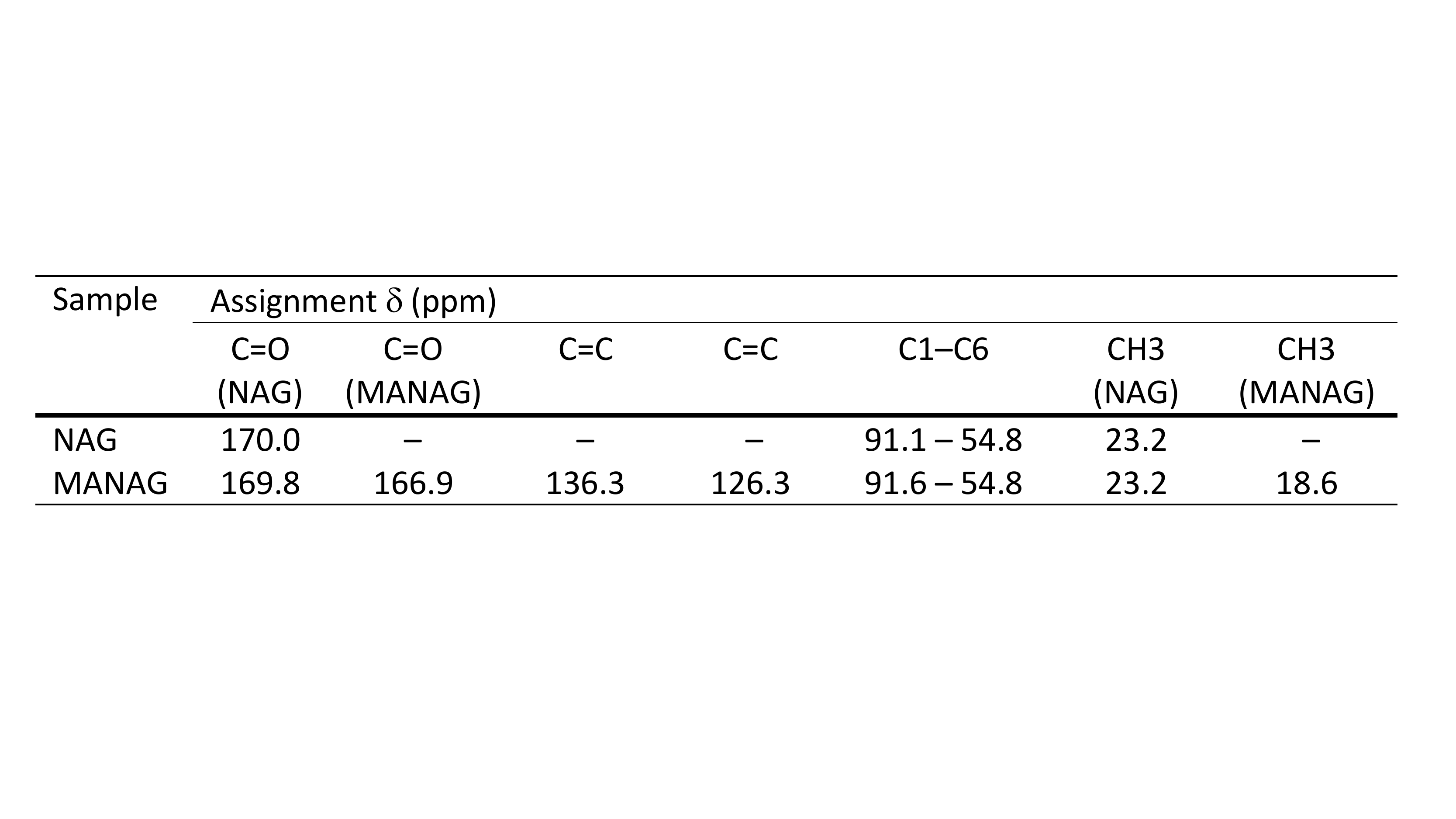}
    \label{fig:NMR}
\end{table}

Additionally, $^{1}$H NMR spectra affirm a successful methacrylation by the appearance of further signals compared to NAG. The signals at 5.98\,ppm and 5.65\,ppm refer to the vinyl protons of the carbon double bond\cite{Khor2011, Wang2013}, while the peak at 1.85\,ppm is assigned to the methyl protons of the methacrylic moiety, respectively\cite{Ramilo1994, Khor2011}. 

Due to some persistent impurities of the applied catalysts within the MANAG product, a detailed determination of the degree of methacrylation was impossible, but elemental analysis measurements of the synthesized MANAG indicate a value of about 1.5. Therefore, the resulting polymer networks after polymerization are probably a mixture of linear polymers in case of MANAG monomers with a degree of methacrylation of 1.0 and a crosslinked polymer network, which consists of MANAG molecules with a degree of methacrylation above 1.0.

MANAG is crosslinkable via two-photon absorption using a DLW-setup as well via one-photon absorption employing a UV lamp ($\lambda$ = 254\,nm and 365\,nm).

The synthesized MANAG is soluble in dimethyl sulfoxide (DMSO), dimethylformamide, and dimethylacetamide, whereas the crosslinked monosacharides are insoluble in common organic solvents.

\subsection{Direct laser writing of NAG-based structures}

\begin{figure}[th]
    \centering
    \includegraphics[width=\textwidth]{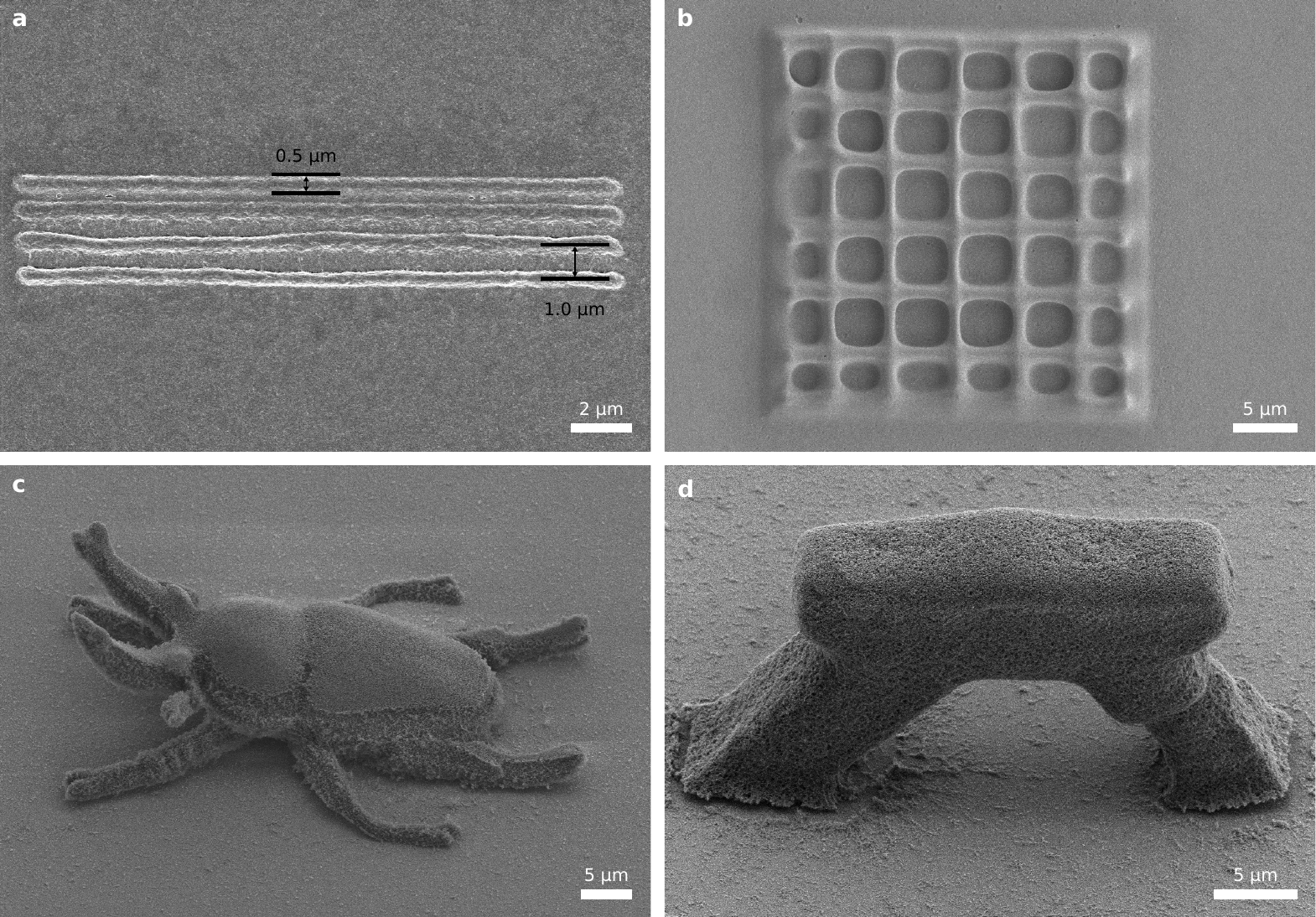}
    \caption{SEM micrographs: (a) printed lines revealing a micron resolution and a sub-micron feature size. The lines are written at a speed of 10 µm/s with 30\% laser power in a resist containing Irgacure 369. (b) Fabricated 2D grid structure at a writing speed of 50\,µm/s and 90\% laser power using Irgacure 819 as photoinitiator. (c) A 3D model of a rainbow stag beetle written at 1000\,µm/s and 90\% laser power in a photoresist containing Irgacure 369. (d) Printed arch at 100\,µm/s and 30\% laser power using Irgacure 819, revealing a free standing structure. The images in (a), (b) show top views on the respective structure while the images in (c), (d) depict side views at 45°. 
    }
    \label{fig:DLW_structures}
\end{figure}

The functionalized NAG is dissolved in DMSO and mixed with different photoinitiators to obtain feasible NAG-based photoresists. In preliminary tests, the commercial photoinitiators Irgacure 369 and Irgacure 819 provide the best structure quality and hence are used for investigating the capability of the NAG-based photoresists. Figure \ref{fig:DLW_structures} shows a selection of different structures, which can be fabricated with NAG-based resists using Irgacure 369 (a,c) or Irgacure 819 (b,d) as photoinitiator.

Fabrication of typical 2D structures such as lines and grids are depicted in Figure \ref{fig:DLW_structures}a,b. The lines in Figure \ref{fig:DLW_structures}a reveal a lateral resolution of about 1\,µm and a feature size of around 500\,nm, which is comparable to the specifications of a previously reported cellulose-based photoresist \cite{Rothammer18}. Achieving feature sizes and a resolution on the scale of about one micron is important for the fabrication of biomimetic structures as shown by the wide range of functional nano- and microstructures found for example in insects \cite{Schroeder18}.

While 2D structures can be successfully printed in NAG-based photoresists, the writing speed is limited for both photoinitiators. To fabricate a 2D grid at a writing speed of 50\,µm/s as displayed in Figure \ref{fig:DLW_structures}b a laser power of already 90\% (see Methods) has to be applied when using Irgacure 819. Similar results are obtained, when Irgacure 369 is used. Since higher writing speeds require higher laser power to achieve the same dose, no well defined 2D structures are obtained at a significantly higher writing speed. 

Besides 2D structures also complex 3D architectures can be fabricated with NAG-based resists as shown in Figure \ref{fig:DLW_structures}c revealing a miniaturized 3D model of a rainbow stag beetle\footnote{The free 3D model of the beetle is obtained from the website www.ameede.net}. Figure \ref{fig:DLW_structures}d depicts a printed arch consisting of two skewed pillars capped by an architrave. This shows that also free standing 3D structures can be realized underlining the capability of the resist for 3D microprinting.

\subsection{Selecting suitable sensitizer-photoinitiator pairs}

As shown in the previous section, the NAG resist can be structured using a single photoinitiator. However, fabricating non-bulky structures requires relatively low writing speeds compared to many other photoresists, which can be printed at several mm/s or even faster \cite{Kiefer20}.

This disadvantage can be reduced, when a proper sensitizer is added to the photoresist. In general, a sensitizer is also a photoinitiator, but it has to fulfill additional demands (as stated below) to form a sensitizer-photoinitiator pair capable to show a sensitization effect. Therefore, choosing an appropriate sensitizer-photoinitiator pair is conceptionally different from mixing two photoinitiators (as reported elsewhere) \cite{Serien17}.

In a sensitizer-photoinitiator pair the sensitizer and the photoinitiator are distinguished by the following two points. First, at the wavelength used the sensitizer should possess a higher absorption than the photoinitiator. In the case of DLW-suitable photoresists the TPCS of the sensitizer should be larger than the TPCS of the photoinitiator. Second, the sensitizer needs a triplet energy, which is higher than that of the photoinitiator. Hence, the energy in the excited triplet state of the sensitizer can be transferred to the triplet state of the photoinitiator initializing the formation of a radical. Meanwhile the sensitizer returns to the ground state being retained for a new cycle \cite{Rist92}.

Since the photoinitiator is responsible for the initiation of the polymerization, it should be capable to effectively crosslink the monomer. This is the case for Irgacure 369 and Irgacure 819 as shown in the previous section, so these compounds can be considered as the photoinitiator part of the sensitizer-photoinitiator pair.

\begin{figure}[th]
    \centering
    \includegraphics[width=\textwidth]{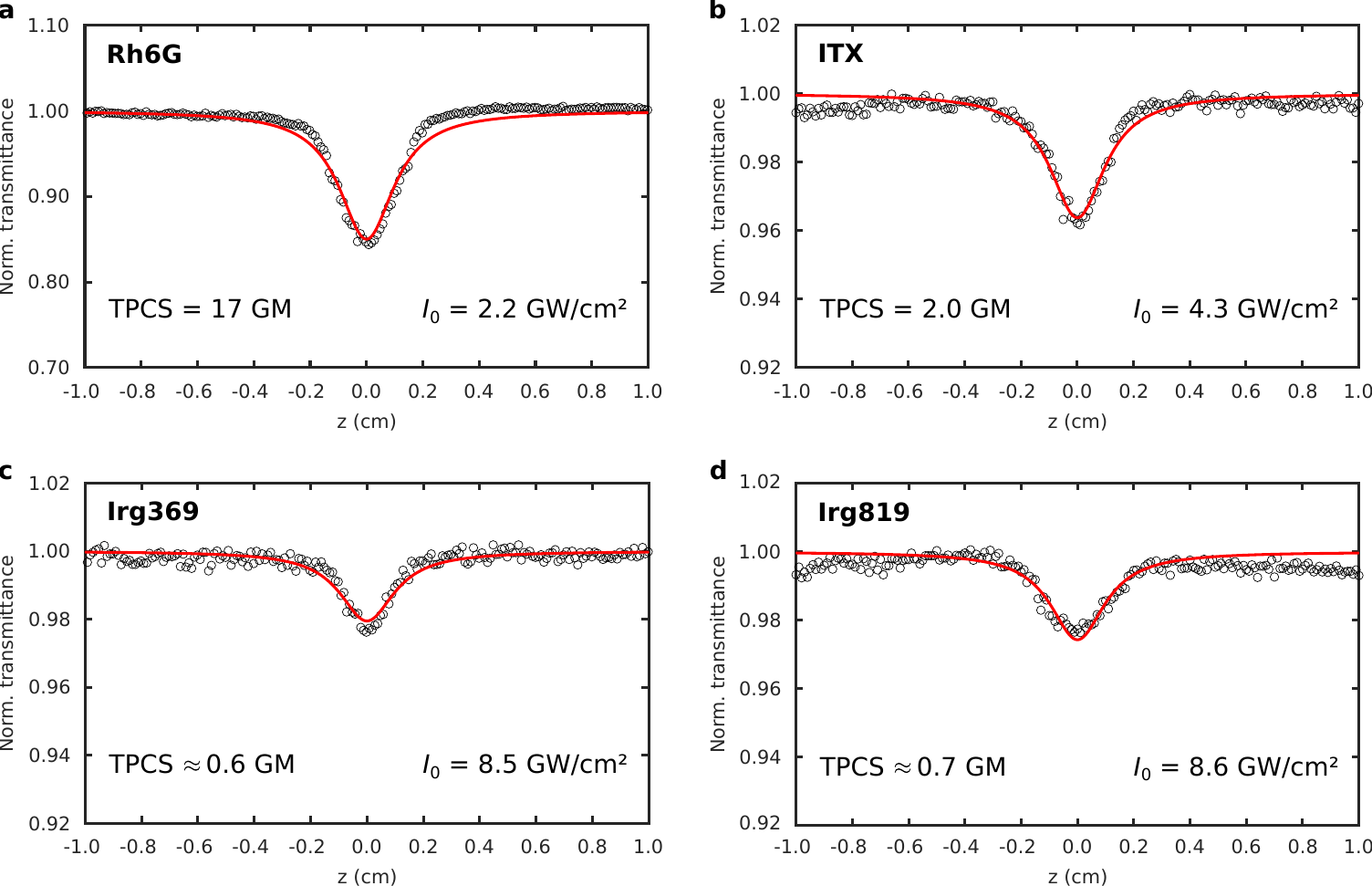}
    \caption{(a)-(d) Open aperture z-scan measurement at 780 nm of (a) Rhodamine 6G in MeOH, (b) ITX in DMSO, (c) Irgacure 369 in DMSO, and (d) Irgacure 819 in DMSO. As stated in the subfigures the intensity at the focus $I_0$ is adjusted to account for different two-photon absorption strengths. Note that the ordinate in (a) is scaled differently from (b)-(d) to improve visibility of the characteristic dip in all cases.
    }
    \label{fig:z-scan}
\end{figure}

To find appropriate candidates for the sensitizer part, the triplet energy is taken into account. Irgacure 369 and Irgacure 819 possess a triplet energy of 251\,kJ/mol and 232\,kJ/mol \cite{Green10}, respectively, thus isopropylthioxanthone (ITX, 257\,kJ/mol \cite{Green10}) and benzophenone (BP, 289\,kJ/mol \cite{Green10}) are identified as possible sensitizers. However, to act as a sensitizer they also have to provide a higher TPCS than the photoinitiators.

The TPCS of the chosen photoinitiators (Irgacure 369, 819) and sensitizers (ITX, BP) are measured using an open aperture z-scan setup (see Methods). In short, a cuvette containing a solution of the respective substance is moved through the focal spot of a laser beam along the axial direction while the transmittance is collected as a function of the scan position. Hence, the laser intensity dependent absorption is recorded yielding direct access to the TPA coefficient \cite{Sheik90,Nag09}. To validate the setup the known TPCS of Rhodamine 6G (Rh6G) in methanol at 780\,nm is used as reference \cite{Nag09}. Measuring the TPCS of Rh6G (see Figure \ref{fig:z-scan}a) reveals a value of 17\,GM (Goeppert Mayer) in good accordance with the literature value of 16\,GM.         

Since the TPCS depends also on the solvent \cite{Nag09}, the photoinitiators and sensitizers of interest are dissolved in DMSO, which is used to mix the photoresists. Measuring at 780 nm (the wavelength applied for DLW) BP does not show any TPA up to the maximum peak intensity of our setup and is therefore not applicable as sensitizer. However, ITX reveals a TPCS of 2\,GM, which is indeed remarkably higher than the TPCS for Irgacure 369 and 819 which are both in the range of 0.6 to 0.7\,GM, as it can be discerned in Figure \ref{fig:z-scan}b-d. Note that the peak intensity for both Irgacure initiators has to be increased by a factor of two compared to ITX to obtain the characteristic dip. This result is in accordance with the expectations since ITX possesses its TPA maximum at 760\,nm while Irgacure 369 and 819 reach their TPA maxima at 670\,nm and\,600 nm, respectively \cite{Schafer04}. After identification of ITX as a suitable sensitizer, two sensitizer-photoinitiator pairs are tested, consisting of either ITX and Irgacure 369 or ITX and Irgacure 819.

\subsection{Sensitization effect for two-photon polymerization}

To investigate the impact of the sensitization effect on two-photon polymerization various photoresists are mixed (see Methods) containing either Irgacure 369, 819 or ITX as well as one of the two sensitizer-photoinitiator pairs mentioned above. For a fair comparison between the different resists the sum of sensitizer and photoinitiator molecules in the resists containing a sensitizer-photoinitiator pair is equal to the amount of photoinitiator molecules in the other resists. In addition only structures written at the same writing condition, i.e. writing speed and laser power, are directly compared to each other.

\begin{figure}[th]
    \centering
    \includegraphics[width=\textwidth]{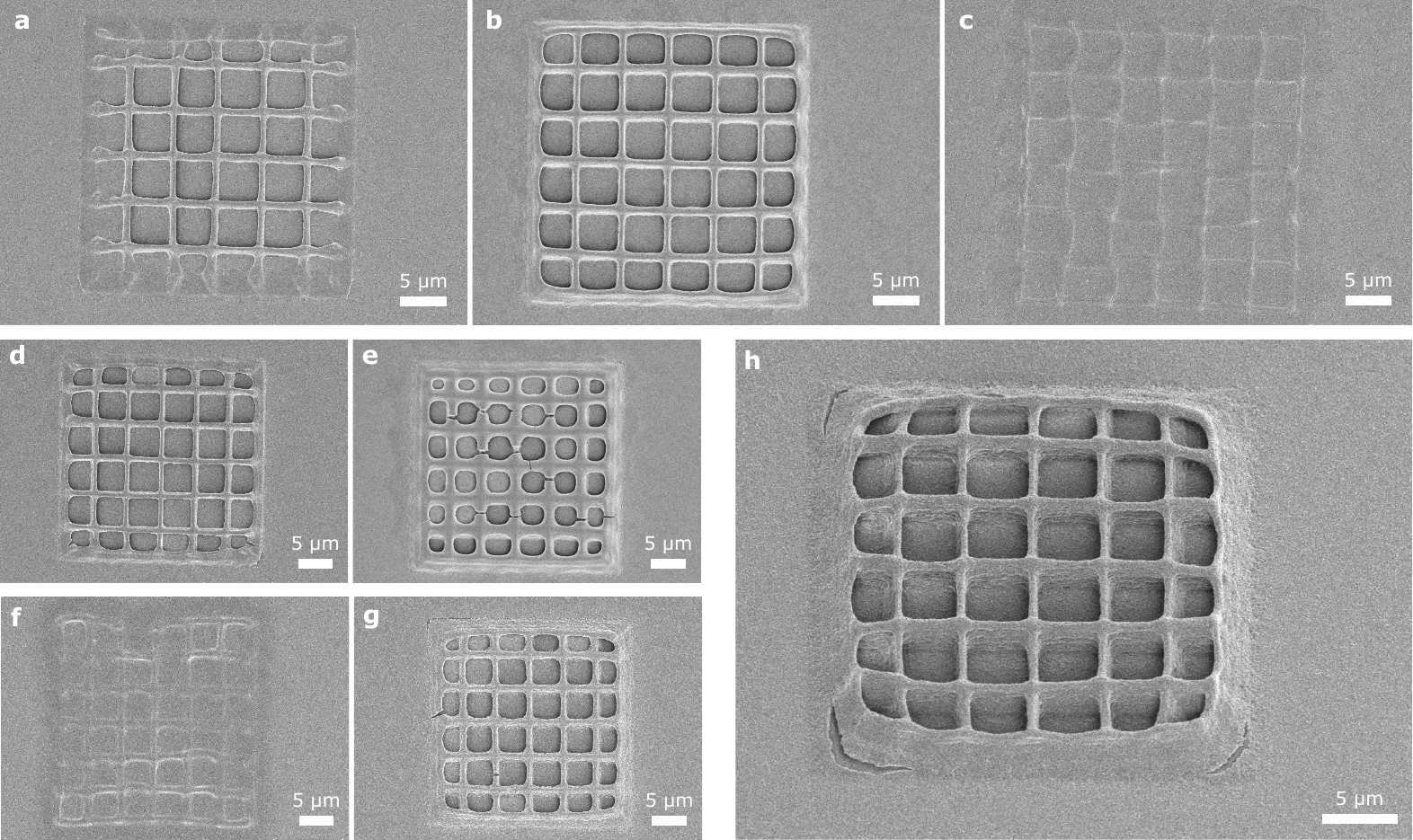}
    \caption{SEM micrographs. (a)-(c) Direct laser written 2D grid structure in a resist containing (a) Irgacure 369, (b) a combination of Irgacure 369 and ITX, and (c) ITX. All grids are written at a speed of 20\,µm/s and 30\% laser power. (d), (e) Fabrication of the same structure using (d) Irgacure 369 and (e) a mixture of Irgacure 369 and ITX at a writing speed of 20\,µm/s and 50\% laser power. (f), (g) Printed grids for a resist containing (f) Irgacure 819 and (g) the combination of Irgacure 819 and ITX. The writing speed is 50 µm/s and a laser power of 40\% is used. (h) Grid fabricated in the same resist at 500 µm/s and 50\% laser power. The SEM image depicts a side view at an angle of 30°.
    }
    \label{fig:sensitization}
\end{figure}

Figure \ref{fig:sensitization}a-c shows a two-dimensional grid of lines written in a resist containing only Irgacure 369 (a), a mixture of Irgacure 369 and ITX with a ratio of 3:1 (b), and pure ITX (c). For Irgacure 369 and the sensitizer-photoinitiator pair raised lines can be clearly seen at a laser power of 30\% and a writing speed of 20 µm/s. In contrast, almost no lines are observed when the sensitizer is used alone under the same writing conditions. As shown by the weak contrast between the lines and the substrate (see Figure \ref{fig:sensitization}c) there is only a tendency of forming lines indicating that higher laser power is needed to form actual solid structures. Indeed, applying laser power close to 100\% it is also possible to achieve solid lines using ITX.

However, comparing Figure \ref{fig:sensitization}a and b reveals that the sensitizer-photoinitiator pair clearly outperforms Irgacure 369 in terms of structure quality, as it is recognizable for example at the edge of the grid where several lines are missing in Figure \ref{fig:sensitization}a. This demonstrates that the combination of a photoinitiator and a sensitizer is highly advantageous over using either solely the photoinitiator or the sensitizer.   

Applying a laser power of 50\% improves the overall quality of the structure fabricated with pure Irgacure 369, Figure \ref{fig:sensitization}d. Contrary, the same laser power results in a considerably overexposure of the lines in case of the sensitizer-photoinitiator pair, Figure \ref{fig:sensitization}e. This clearly shows that the same dose leads to stronger polymerization, which can be explained by an increased light utilization due to the sensitization effect. 

This effect is not limited to one certain photoinitiator-sensitizer pair and writing condition. Figure \ref{fig:sensitization}f,g shows a grid structure fabricated with only Irgacure 819 and the corresponding sensitizer-photoinitiator pair, respectively. Here, a writing speed of 50\,µm/s and a laser power of 40\% is used. At these conditions no raised lines are observable using solely Irgacure 819, Figure \ref{fig:sensitization}f. However, for a laser power of 90\% it is also possible to write solid structures using Irgacure 819 as shown in Figure \ref{fig:DLW_structures}b. On the other hand the sensitizer-photoinitiator pair enables structure fabrication already at a 40\% laser power where neither Irgacure 819 nor ITX alone yield similar results, Figure \ref{fig:sensitization}g. 

It should be noted that the ratio between Irgacure 819 and ITX is 7:1 for the presented resist, delivering a further proof that the sensitization effect also works for two-photon polymerization. Since the sensitizer just transfers the absorbed energy without being destroyed in general, it is expected that the sensitization effect occurs already for a small addition of sensitizer.

While the maximum writing speed for non-bulky structures is limited to $\approx 50$ µm/s using a single photoinitiator, applying the sensitization effect indeed greatly enhances the writing speed. Figure \ref{fig:sensitization}h shows a grid printed at 500\,µm/s and 50\% laser power using the resist containing a mixture of Irgacure 819 and ITX. Thus, the writing speed can be increased by an order of magnitude for the investigated NAG resist underlining the benefit of the sensitization effect for two-photon polymerization and DLW.

\section{Conclusions}

We have synthesized a novel functionalized derivative of N\hbox{-}acetyl\hbox{-}\textsc{D}\hbox{-}glucosamine, which enables the fabrication of three-dimensional micro-architectures using DLW. Thus, the range of available bio-sourced monomers is expanded to those of animal origin having potential for a closer biomimicry of animal micro-structures. In addition, the successful establishment of a NAG-based photoresist is an important step towards developing novel resists utilizing more complex animal-based biopolymers such as chitin.

While the methacrylated NAG can be crosslinked using a single photoinitiator, it is demonstrated that the addition of a sensitizer substantially improves the quality of the printed structure and enables remarkably higher fabrication speed. This underlines that the sensitization effect known from pigmented UV curing can be successfully transferred from one-photon to two-photon absorption processes for the first time and hence become beneficial for DLW. Moreover, the effect is not limited to one specific sensitizer-photoinitiator pair indicating that sensitization is an universal effect also in case of two-photon absorption. Therefore, many other sensitizer-photoinitiator pairs might be found possibly outperforming the presented pairs or even enabling the use of photoinitiators, which are currently inaccessible due to a diminishing two-photon absorption cross section at the wavelengths typically applied in commercial DLW systems.


\section{Experimental Section}

\subsection{Synthesis of the photo-crosslinkable NAG}
3.05\,g of NAG (Alfa Aesar, Germany) was dissolved in 100\,ml dimethylformamide ($\geq$98\%, VWR Chemicals, Germany) at 58\,°C under permanent stirring and reflux conditions. After the solution was allowed to cool to RT, 5.8\,ml methacrylic acid anhydride (94\%, Sigma-Aldrich, Germany), 0.1\,g 4\hbox{-}dimethylaminopyridine (DMAP, 99\%, Acros Organics, USA) and 4.04\,g \textit{N,N}\hbox{-}dicyclohexylcarbodiimide (DCC, 99\%, Alfa Aesar, Germany) were added. The reaction was performed under permanent stirring for 45\,h. Subsequently, the reaction mixture was concentrated by employing a rotary evaporator. The methacrylated NAG was then precipitated in an excess of diethyl ether (technical, VWR Chemicals, Germany) within an ice bath. The product was washed five times with diethyl ether using a centrifuge at 4000\,rpm at 20\,°C for 20\,min. Finally the methacrylated NAG was obtained after evaporating the solvent by employing a rotary evaporator. All chemicals were used as received without further purification.

\subsection{UV curing}
For the preparation of the photoresist, 50.0\,mg MANAG were dissolved in 250\,µL dimethyl sulfoxide. Subsequently, 1\,mg Irgacure 369 was added under stirring for 30\, minutes. Then, the photoresist was irradiated simultaneously with wavelengths of 254\,nm and 365\,nm for 30\,minutes, whereas the sample was placed 5\,cm away from the UV light source. UV induced crosslinking was performed with a 8\,Watt UV lamp (Herolab, Germany).

\subsection{Fourier transform infrared spectroscopy}
Fourier transform infrared (FTIR) spectra were recorded on the powder of NAG and functionalized NAG using a Frontier MIR spectrometer (L1280018) performed with an attenuated total reflection (ATR) diamond (PerkinElmer, Germany). The spectra were recorded between 4000 and 400\,cm$^{-1}$ with a resolution of 4\,cm$^{-1}$ and at least 8\,scans.

\subsection{NMR spectroscopy}
NMR spectra were acquired on a Jeol ECS-400 NMR (Japan). The evaluation was performed by Delta v 5.0.4 software (Jeol). Samples were analyzed in dimethyl sulfoxide\hbox{-}d$_6$ solutions at 25\,°C. $^{13}$C NMR spectra were recorded with 2048\,scans, whereas $^{1}$H NMR spectra were measured with 32\,scans.

\subsection{Elemental analysis}
The elemental composition of the unmodified and modified NAG was measured by a EuroEA-Elemental Analyser (Eurovector, Italy). 1–3\,mg powder in tin crucibles was analyzed. This enabled the determination of the degree of methacrylation of NAG.

\subsection{Z-scan measurements}
For open-aperture z-scan measurements the beam of a femtosecond Ti:Sa laser (Chameleon, Coherent Inc., USA) is focused with a 20\,cm focal lens onto a point within the travel range of a motorized stage (Soloist, Aerotech GmbH, Germany). About 10\,cm behind the focus the light is collected with a second lens focusing the light on a detector (DET36A2, Thorlabs Inc., USA). The output of the detector is fed into a lock-in amplifier (SR830 DSP, Standford Research Systems, USA). Thereby, the lock-in amplifier is triggered by an optical chopper (MC2000, Thorlabs Inc.) equipped with a chopper blade with 50\% duty cycle, which is placed in the beam path right in front of the first lens. The output signal of the lock-in amplifier as well as the current stage position are sent to a computer and recorded by a self-written LabVIEW program (National Instruments, USA). To measure the TPCS of the photoinitiators they are dissolved in DMSO (Sigma-Aldrich) in a concentration of 0.05\,M, filled into a cuvette with 1\,mm optical path length (Hellma GmbH \& Co. KG, Germany) and scanned through the focus of the laser beam using the stage. In addition Rhodamine 6G (Acros Organics) dissolved in the same concentration in methanol (VWR International GmbH, Germany) is used as a reference. The TPA coefficient $\beta$ can be obtained from the measured normalized transmittance curve by fitting with the equation $T = 1 - \frac{\beta I_{0} L}{2^{3/2}(1 + (z^2/z_{0}^{2}))}$, where $I_0$ is the peak intensity at the focus, $L$ the sample thickness and $z_0$ the diffraction length. Except for $\beta$ all other quantities depend only on the setup and are accessible as explained in more detail elsewhere \cite{Sheik90,Nag09}. The TPCS is directly related to $\beta$ via $\text{TPCS} = \frac{\beta h \nu \times 10^3}{N_{\rm A} c}$ with frequency $\nu$ of the used laser, the Avogadro constant $N_{\rm A}$ and the concentration of photoinitiator $c$ \cite{Nag09}. All results are given in the common unit Goeppert Mayer (GM).          

\subsection{Preparation of photoresists}
Prior to mixing the photoresists different stock solutions are prepared. The stock solutions contain DMSO (Sigma-Aldrich) and either Irgacure 369, Irgacure 819 (both TCI chemicals, Japan) or ITX (Sigma-Aldrich) each in a concentration of 0.05\,M. All stock solutions are stirred until the photoinitiator is completely dissolved. For mixing the photoresists first the required amount of functionalized NAG is weighed. Subsequently, stock solution is added until a concentration of 250\,g NAG per liter stock solution is reached. For resists containing only one photoinitiator the stock solutions are added as prepared. For the both sensitizer-photoinitiator pairs the stock solutions are first mixed in a ratio of 3:1 (Irgacure 369 to ITX solution) as well as 7:1 (Irgacure 819 to ITX solution) before adding the mixture to the NAG. Afterwards the resists are stirred over night using a magnetic stirrer. To avoid the inclusion of unsolved particles the resists are finally transferred to a reaction vial and separated for 20\,minutes in a centrifuge (Mini-Zentrifuge, Carl Roth GmbH + Co. KG, Germany).

\subsection{Direct laser writing}
All structures are printed with a Photonic Professional GT (Nanoscribe GmbH, Germany) operating at a wavelength of 780\,nm with a laser power of 57.8\,mW (= 100\%). The photoresists are extracted from the vial using a syringe and a drop of resist is placed on top of a 170\,µm thick glass substrate (Gerhard Menzel GmbH, Germany). The top side of the substrate is coated with a thin layer of aluminum oxide using atomic layer deposition (R-200 Standard, Picosun Oy., Finland) to enable easier detection of the glass-resist interface. The used sample holder allows for placing an uncoated coverslip slightly above the substrate sandwiching the resist in between. This is done because DMSO is hygroscopic with the detrimental effect of blurring the resist if it is exposed to air too long. During writing the laser beam is focused into the resist through the substrate, using a drop of immersion oil (Immersol 518F, Carl Zeiss AG, Germany) between the objective (63$\times$, NA = 1.4) and the bottom of the substrate. After printing the samples are developed for 90 min in DMSO and subsequently for 10\,min in acetone. For 2D grids and lines the samples are carefully blown dry with a nitrogen pistol. The printed 3D structures are dried using a critical point dryer (EM CPD300, Leica Microsystems GmbH, Germany).

\medskip
\textbf{Supporting Information} \par 
Supporting Information is available from the authors upon request.

\medskip
\textbf{Acknowledgements} \par
D.T.M. and M.R. contributed equally to this work. We gratefully acknowledge financial support from the German Research Foundation DFG within the priority program "Tailored Disorder -  A science- and engineering-based approach to materials design for advanced photonic applications" (SPP 1839). We thank the team of the Nano Structuring Center (NSC) at the Technische Universität Kaiserslautern for their support with scanning electron microscopy.

\medskip
\textbf{Conflict of Interest} \par
The authors declare no conflict of interest.


\newpage
\textbf{Table of Content} \par

\begin{figure}[H]
    \centering    \includegraphics{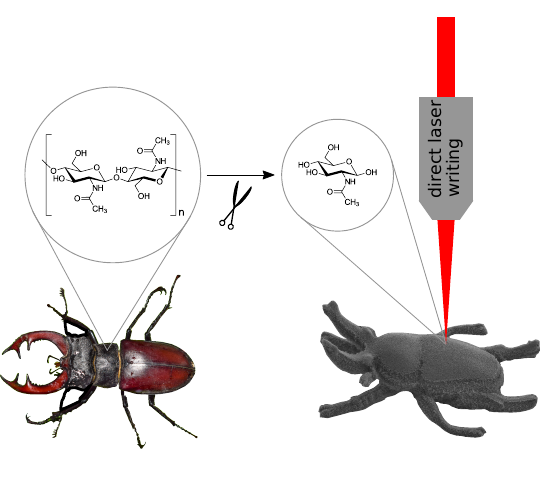}
\end{figure}

A new direct laser writing (DLW) suitable photoresist is presented, which is
based on the chitin monomer N\hbox{-}acetyl\hbox{-}\textsc{D}\hbox{-}glucosamine. Hence, the range of bio-based photoresists is extended to those of animal origin. In addition it is shown that the sensitization effect previously only known for UV curing can also be applied for DLW significantly increasing the achievable writing speed.

\end{document}